%% file: paper_writing.tex
\documentclass[sigconf]{acmart}

\usepackage{algorithmic}
\usepackage{algorithm}
\usepackage{graphicx}
\usepackage{textcomp}
\usepackage{threeparttable}
\usepackage{pifont}
\usepackage{booktabs}
\usepackage{url}
\usepackage{nicefrac}
\usepackage{multirow}
\usepackage{hyperref}
\usepackage{tikz}
\usepackage{dblfloatfix}
\usepackage{soul}
\sethlcolor{yellow}
\usepackage[table]{xcolor}
\usepackage{colortbl}

\definecolor{1}{RGB}{220,235,255}  
\definecolor{2}{RGB}{255,235,220}    

\newcommand*\circled[1]{\tikz[baseline=(char.base)]{
  \node[shape=circle,draw,inner sep=0.5pt] (char) {#1};
}}

\AtBeginDocument{%
  }

\definecolor{lightgreen}{RGB}{0,180,0}
\definecolor{lightblue}{RGB}{150,210,255}

\copyrightyear{2026}
\acmYear{2026}
\setcopyright{cc}
\setcctype{by}
\acmConference[ICCAD '26]{IEEE/ACM International Conference on Computer-Aided Design}{November 08--12, 2026}{San Jose, CA, USA}
\acmBooktitle{IEEE/ACM International Conference on Computer-Aided Design (ICCAD '26), November 08--12, 2026, San Jose, CA, USA}
\acmDOI{10.1145/3831252.3833983}
\acmISBN{979-8-4007-2873-0/2026/11}

\begin{document}

\title{Lonic: Algorithm-Hardware Co-Design for Energy-Efficient Fully Local Online SNN Training with INT4 Precision}

\author{Peilin Chen}
\email{peilin@virginia.edu}
\affiliation{
  \institution{University of Virginia}
  \city{Charlottesville}
  \state{VA}
  \country{USA}
}

\author{Xiaoxuan Yang}
\email{xiaoxuan@virginia.edu}
\affiliation{
  \institution{University of Virginia}
  \city{Charlottesville}
  \state{VA}
  \country{USA}
}

\input{sec0_abs}

\ccsdesc[500]{Hardware~Application-specific VLSI designs}

\keywords{Spiking neural networks, Fully local online learning, Energy efficiency, Low-precision training}

\maketitle

\input{sec1_intro}

\input{sec2_background}

\input{sec3_method}

\input{sec4_experiment}

\input{sec5_conclusion}

\bibliographystyle{ACM-Reference-Format}
\bibliography{citations}

\end{document}

%% file: sec0_abs.tex
\begin{abstract}
Spiking neural networks~(SNNs) have recently attracted increasing attention as an energy-efficient learning paradigm. Existing works also propose temporally and fully local online SNN training algorithms to address memory and computation overhead. However, they do not consider whether the algorithmic advantages can be effectively translated into real-device efficiency. To address this challenge, we present Lonic, an algorithm-hardware co-design for energy-efficient and scalable fully local online supervised SNN learning. \textit{On the algorithm side,} we implement an INT4 low-precision training algorithm for fully local online SNN learning while maintaining accuracy. \textit{On the hardware side,} to leverage the benefits of the proposed algorithm, we introduce reconfigurable multiplier-free integer PE arrays, dual-optimization zero-gating strategy, temporal prefix-accelerated local learning dataflow, and low-precision weight movement to significantly improve training efficiency. Compared to Apple M4 and Nvidia V100 GPUs, Lonic achieves average energy efficiency improvements of 17.44x and 66.28x, respectively, along with speedups of 3.25x and 1.02x, respectively. Moreover, Lonic achieves 15.95x~(14.64x) and 1.52x~($7.28$x) energy efficiency~(area efficiency) over ASIC TPU-like and H2Learn accelerators, respectively. The code for Lonic is available at \url{https://github.com/peilin-chen/Lonic}.
\end{abstract}

%% file: sec1_intro.tex
\section{Introduction}
\label{section_1}

Spiking neural networks~(SNNs) have emerged as a promising paradigm for energy-efficient machine learning due to their event-driven computation and biological plausibility~\cite{eshraghian2023training}. Unlike artificial neural networks~(ANNs) that rely on continuous-valued activations, SNNs transmit information via binary spikes encoded over time, enabling high spatial/temporal sparsity and low-precision computation~\cite{eshraghian2023training, jiang2024ndot, yin2024loas,haoxuan2025neuromorphic}. To achieve competitive accuracy as ANNs, most SNN training frameworks adopt the backpropagation through time~(BPTT) algorithm with surrogate gradient~\cite{xiao2022online, zheng2021going, lecun2002gradient, chen2026spikint}. However, BPTT is incompatible with biological online learning and requires storing past inputs and states in memory~\cite{jiang2024ndot, xiao2022online, meng2023towards, eshraghian2023training}. As a result, BPTT's memory usage is proportional to the product of time and the number of neurons~\cite{eshraghian2023training, xiao2022online}.

\begin{figure}[tb]
    \centering
    \setlength{\abovecaptionskip}{0pt}
    \includegraphics[width=0.87\linewidth]{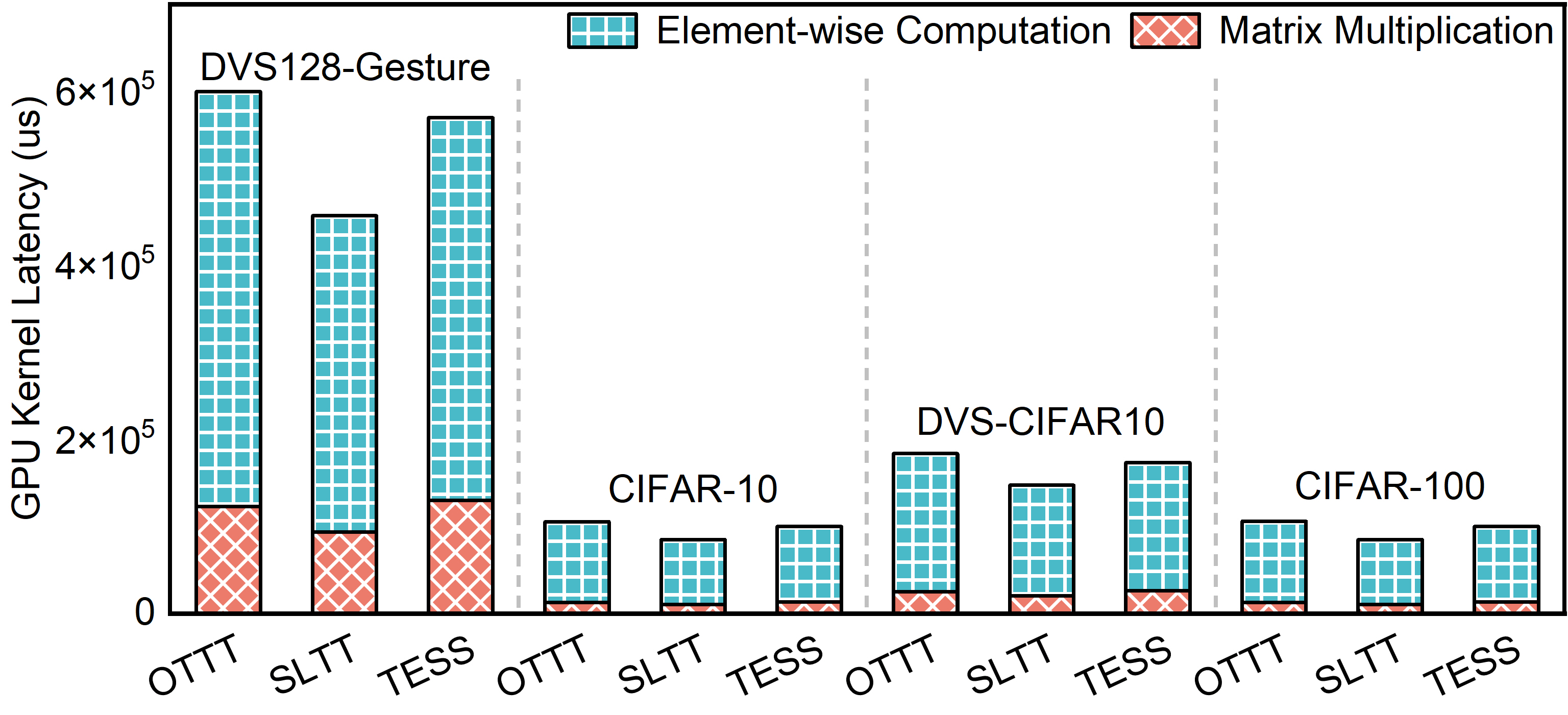}
    \caption{Runtime profile of OTTT, SLTT, and TESS under Nvidia GPU across four datasets~(training batch size=1).}
    \Description{fig1}
    \label{fig1}
    \vspace{-10pt}
\end{figure}

Prior work proposes various online SNN learning algorithms to address these problems~\cite{xiao2022online, meng2023towards, apolinario2025tess}. OTTT~\cite{xiao2022online} enables forward-in-time learning by tracking presynaptic activities and leveraging instantaneous gradients. SLTT~\cite{meng2023towards} claims that temporal gradients are unimportant and should be abandoned to reduce time complexity. Different from OTTT and SLTT, which rely on the global error backpropagation~(temporally local learning), TESS~\cite{apolinario2025tess} proposes a scalable temporally and spatially local SNN learning rule to address both memory and computation overhead while maintaining comparable accuracy. However, TESS~\cite{apolinario2025tess} only demonstrates theoretical reductions in MAC operations~(205x-661x) and does not explore whether these benefits translate into practical hardware efficiency. To understand the real-device efficiency of the fully local learning rule, we profile the training latency of OTTT, SLTT, and TESS on the VGG11 network using Nvidia GPU~(Fig.~\ref{fig1}). We can observe that element-wise computations dominate the overall latency, as the small matrix multiplications under the online learning rule are efficiently handled by the GPU. Moreover, the theoretical MAC reduction of TESS does not translate into hardware-level speedup. 

\begin{figure*}[tb]
    \centering
    \setlength{\abovecaptionskip}{0pt}
    \includegraphics[width=\linewidth]{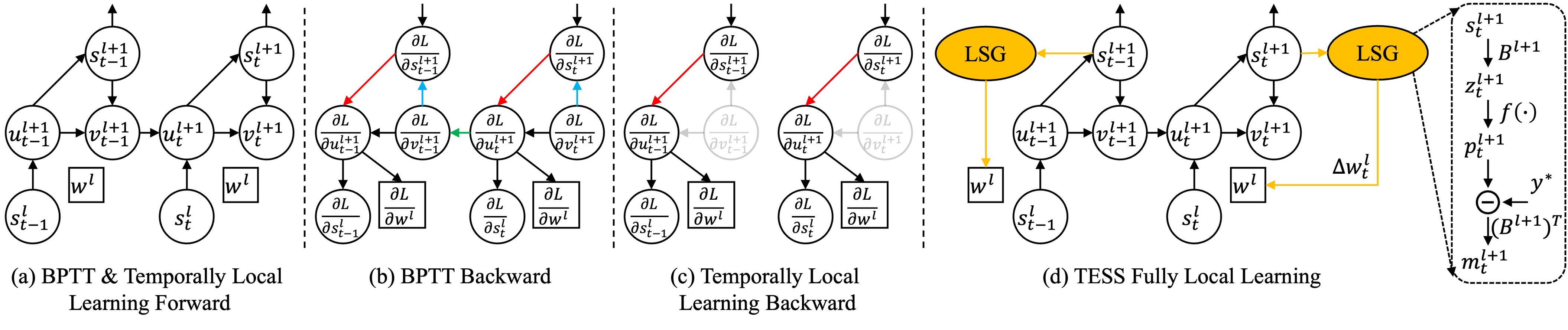}
    \caption{Comparison of three training algorithms of SNNs: BPTT, temporally local online learning, and fully local online learning. $s^l_t$, $u^l_t$, and $v^l_t$ denote the spike output, membrane potential after charging, and membrane potential after resetting of layer $l$ at timestep $t$. $L$ and $w^l$ represent the loss and weights of SNN in layer $l$. LSG represents the learning signal generator.}
    \Description{fig2}
    \label{fig2}
\end{figure*}

Despite its success on large networks and datasets, the fully local learning algorithm TESS rarely considers hardware efficiency, making the theoretical computation reduction difficult to realize on real hardware. To bridge this gap, we propose Lonic, an algorithm-hardware co-design for energy-efficient and low-latency local online SNN learning. \textbf{At the algorithm level,} Lonic enables INT4 low-precision training for SNNs based on the fully local learning paradigm without accuracy degradation, allowing matrix multiplications to be computed using integer arithmetic and facilitating low-overhead hardware implementation. \textbf{At the hardware level,} Lonic proposes a specialized training accelerator that translates algorithmic advantages introduced by low-precision training into practical hardware efficiency via multiplier-free integer PEs, increased weight sparsity, reduced off-chip data movement, etc. Specifically, the contributions of this work are as follows:
\begin{enumerate}
    \item To the best of our knowledge, Lonic is the first work to implement an INT4 low-precision training algorithm for fully local online SNN learning while maintaining accuracy.
    \item We design three types of multiplier-free PE arrays, including reconfigurable PE supporting INT4/8/16, binary-ternary PE, and ternary-INT8/16 PE, to efficiently execute matrix multiplication using integer arithmetic.
    \item We propose a dual-optimization zero-gating strategy that leverages spike sparsity and weight sparsity introduced by low-precision training flow to reduce energy and improve throughput simultaneously. 
    \item We introduce a temporal prefix-accelerated local learning dataflow to enable simultaneous forward, backward, and weight update across all timesteps for single SNN layer.
    \item Benefit from the low-precision training algorithm, Lonic transfers only weights with INT4 bit-width between the accelerator and off-chip memory to reduce data movement.
\end{enumerate}

%% file: sec2_background.tex
\section{Background}
\label{section_2}

\subsection{Online Learning Algorithm for SNNs}

As illustrated in Fig.~\ref{fig2}, SNN training algorithms can be categorized based on their temporal and spatial dependencies in gradient propagation, ranging from BPTT to fully local online learning. These three learning rules share the same forward process~(Fig.~\ref{fig2}(a) and (d)) based on the popular leaky integrate-and-fire~(LIF) neuron~\cite{fang2023spikingjelly, lobo2020spiking}, which can be formulated as follows:
\begin{align}
u^{l+1}_t &= \beta v^{l+1}_{t-1}+w^ls^l_t \label{equation1}, \\
s^{l+1}_t &= \Theta(u^{l+1}_t - \theta) \label{equation2}, \\
v^{l+1}_t &= u^{l+1}_t-\theta s^{l+1}_t \label{equation3},
\end{align}
where $\beta$ is the leaky constant, modeling the decay of potential at each timestep. $\Theta(x)$ denotes the Heaviside step function. $\theta$ represents the firing threshold. 

However, BPTT with surrogate gradients~(SG)~\cite{neftci2019surrogate} propagates gradients along both temporal and spatial dimensions~(Fig.~\ref{fig2}(b)), which suffers from significant memory overhead and is inconsistent with biological online learning~\cite{xiao2022online}. Different from BPTT, temporally local online learning, such as OTTT~\cite{xiao2022online} and SLTT~\cite{meng2023towards}, ignores the temporal dependency during error backpropagation and enables forward-in-time training, as shown in Fig.~\ref{fig2}(c). To further reduce computational overhead, fully local online learning~\cite{apolinario2025tess} relies only on locally available signals for spatial gradient rather than global backpropagation across layers~(Fig.~\ref{fig2}(d)). The weight gradient of this process can be formulated as follows:
\begin{align}
\nabla w_t^l &=
\left( m_t^{l+1} {\odot} \alpha_{pre} \Psi\!\left(u_t^{l+1}\right) \right) {\otimes} q_t^{l+1}
{+}
\left( m_t^{l+1} {\odot} \alpha_{post} h_t^{l+1} \right) {\otimes} s_t^l \label{equation4}, \\
q_t^{l+1} &= \lambda_{pre}q_{t-1}^{l+1}+s^l_t \label{equation5}, \\
h_t^{l+1} &= \lambda_{post}h^{l+1}_{t-1}+\Psi\!\left(u_{t-1}^{l+1}\right) \label{equation6},
\end{align}
where $\odot$ and $\otimes$ represent the element-wise and outer product, respectively. $q_t^{l+1}$ and $h_t^{l+1}$ are the recurrent variables. $\alpha_{pre}$ and $\lambda_{pre}$ control the magnitude and decay of the causal term, whereas $\alpha_{post}$ and $\lambda_{post}$ characterize the non-causal term~\cite{apolinario2023s, gerstner2018eligibility}. $\Psi\!\left(\cdot\right)$ is an activation function that plays a role similar to SG. Moreover, the right side of Fig.~\ref{fig2}(d) demonstrates the generation of local learning signal $m_t^{l+1}$, which can be formulated by the following equation:
\begin{equation}
m^{l+1}_t = (B^{l+1})^T \left( f(B^{l+1}s^{l+1}_t) - y^* \right) \label{equation7},
\end{equation}
where \( B^{l+1} \) is the projection matrix~(or fixed ternary matrix)~\cite{apolinario2025lls} that captures task-relevant information for the layer, $f(\cdot)$ is the softmax function, and \( y^* \) denotes the labels. These learning methods rely on floating-point~(FP) matrix multiplications, and we aim to achieve energy-efficient low-precision training for fully local online SNN learning.

\begin{figure*}[tb]
    \centering
    \setlength{\abovecaptionskip}{0pt}
    \includegraphics[width=0.85\linewidth]{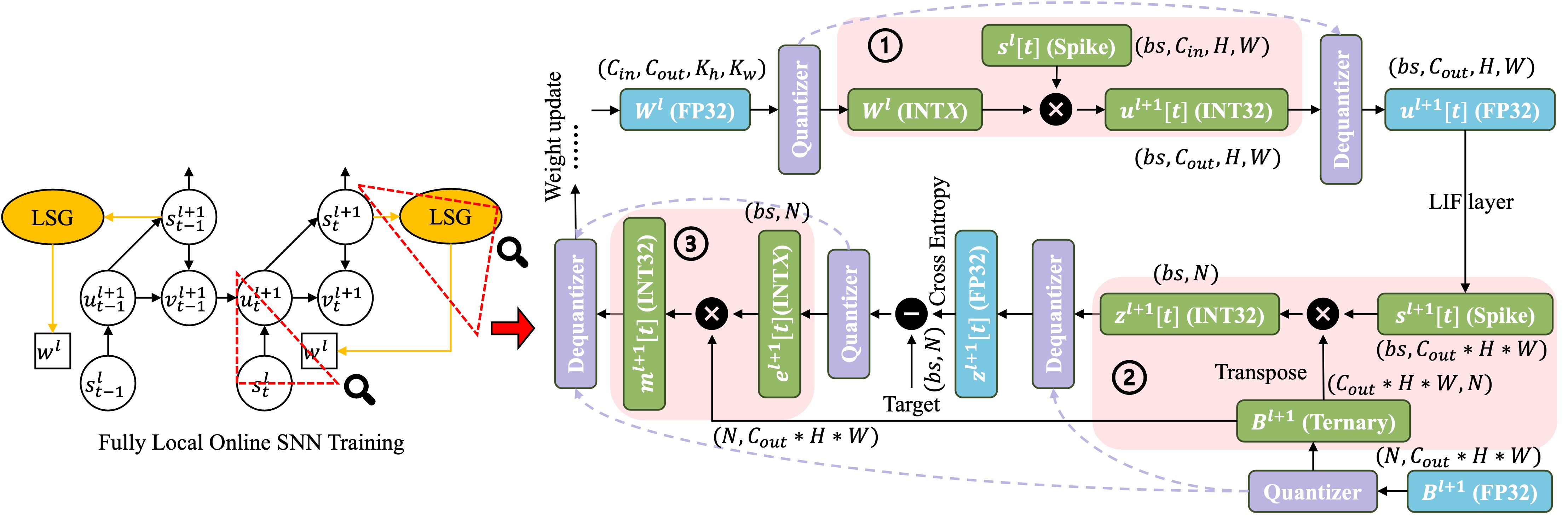}
    \caption{Overview of the low-precision training flow under the fully local online SNN learning paradigm. $C_{in}$/$C_{out}$, $K_h$/$K_w$, $bs$, $H$/$W$, and $N$ represent input/output channel size, kernel height/width, batch size, feature map height/width, and class number.}
    \Description{fig3}
    \label{fig3}
    \vspace{-8pt}
\end{figure*}

\subsection{Scaled Weight Standardization in SNNs}
\label{section_2_2}
Batch normalization~(BN)~\cite{ioffe2015batch, qiao2019micro} is widely adopted in deep learning to stabilize network training. However, BN relies on batch-level statistics, making it unsuitable for online SNN learning, where a single training sample is provided at a time~(batch size=1)~\cite{lobo2020spiking}. To solve this issue, prior works such as OTTT~\cite{xiao2022online}, SLTT~\cite{meng2023towards}, and TESS~\cite{apolinario2025tess} replace BN with scaled weight standardization~(sWS) to eliminate the dependency on large batch sizes~\cite{qiao2019micro}. Specifically, the sWS can be formulated as follows:
\begin{equation}
\hat{w}_{i,j} = \gamma \cdot 
\frac{w_{i,j} - \mu_{w_{i, .}}}
{\sigma_{w_{i, .}} \sqrt{N}} \label{equation8}, 
\end{equation}
where $w_{i,j}$, $\hat{w}_{i,j}$, $N$, $\gamma$, $\mu_{w_{i, .}}$, and $\sigma_{w_{i, .}}$ denote the original weight, standardized weight, fan-in, scaling factor, and the mean and variance of weights, respectively. Although sWS for online SNN learning achieves comparable performance to BN with large batch sizes~\cite{xiao2022online}, it introduces substantial hardware-unfriendly element-wise computations, which dominate the overall latency. as shown in Fig.~\ref{fig1}. Therefore, we need to replace this method with hardware-friendly computations to achieve efficient fully local online SNN learning.

\subsection{Conventional PTQ and QAT}

Post-training quantization~(PTQ) quantizes a pretrained floating-point model without retraining~\cite{han2015deep,yang2022hero,mei2026s}. Quantization-aware training~(QAT) incorporates a \textit{fake quantization} module~(quantization and dequantization operations) during training to simulate quantization effects~\cite{zhang2025survey}. Therefore, computations are still performed in FP for QAT. Despite their effectiveness, PTQ and QAT are primarily designed to improve inference performance rather than reduce training cost. In fact, QAT even introduces additional training overhead due to the \textit{fake quantization} operation, making it more computationally expensive than standard FP training. To improve the training efficiency of fully local online SNN learning, we propose a novel INT4 low-precision training flow that is distinct from PTQ and QAT, while preserving the accuracy~(Section~\ref{section_3}).

\subsection{Related SNN Training Accelerator}

Most existing works~\cite{wei2025prosperity, yin2024loas, mao2024stellar} focus on improving the efficiency of SNN inference by exploiting the inherent sparsity. For example, Prosperity~\cite{wei2025prosperity} proposes \textit{product sparsity} that uses the combinatorial similarity feature in SNN activations to reduce computations. In contrast, hardware support for SNN training is limited, particularly for the online SNN learning paradigm. H2Learn~\cite{liang2021h2learn} targets at reducing the overhead of BPTT-based SNN training. Although one recent study~\cite{siddique2023low} implements specialized hardware for its proposed BP-based online SNN learning algorithm, it is constrained to the small-scale dataset MNIST. To the best of our knowledge, Lonic is the first accelerator to enable energy-efficient and scalable fully local online supervised training for SNNs.

%% file: sec3_method.tex
\section{Lonic Algorithm Design}
\label{section_3}

\subsection{Low Precision Training Algorithm for SNNs}

Fully local SNN learning enables training signals to be computed locally and independently within each layer at each timestep~(Fig.~\ref{fig2}(d)), making it particularly suitable for low-precision training. Since gradients are generated locally without long-range dependency, the introduced quantization errors do not accumulate across timesteps and layers. Based on this observation, we devise a novel low-precision training flow to reduce the computational overhead of general matrix multiplication~(GEMM).

As illustrated in Fig.~\ref{fig3}, fully local learning involves three types of GEMM operations across forward and backward passes: \circled{1} $W^ls^l_t$ for forward computation, and \circled{2} $s^{l+1}_t(B^{l+1})^T$ and \circled{3} $e^{l+1}_tB^{l+1}$ for learning signal generation. The key idea behind our low-precision training flow is to quantize the FP inputs of GEMM, perform integer-based GEMM, and then dequantize the results back to FP. This design trades off quantization error for significantly reduced training cost, unlike QAT, which uses \textit{fake quantization} and still relies on FP computation. Specifically, we adopt the hardware-friendly per-output-channel symmetric quantization~\cite{zhao2020linear} in our training flow. GEMM \circled{1} in Fig.~\ref{fig3} can be formulated as follows:
\begin{align}
W^l(INTX) &= clip(round(\frac{W^l(FP32)}{f_1}),-2^{X-1},2^{X-1}) \label{equation9},\\
u_t^{l+1}(FP32) &= f_1 \times(\mathrm{Conv}(W^l(INTX),s^l_t)) \label{equation10},\\
f_1 &= \frac{max(|W^l(FP32)_{max}|, |W^l(FP32)_{min}|)}{2^{X-1}} \label{equation11},
\end{align}
where $f_1$ and $X$ denote the scale factor and integer bit-width, respectively. Our experiments show that the INT4 precision for GEMM \circled{1} is sufficient to maintain training accuracy in some cases~(Section~\ref{section_5_2}). Different from GEMM \circled{1}, GEMM \circled{2} and \circled{3} incur negligible quantization error, as $B^{l+1}$ is a ternary matrix with values in \{$-\epsilon$, $0$, $+\epsilon$\}, where $\epsilon$ is a FP data, making its quantization inherently lossless. Although quantization of $e^{l+1}_t(FP32)$ in GEMM \circled{3} introduces error, it is negligible in practice at INT8/16 precision due to the small batch size~(bs=1) and limited class number. GEMM \circled{2} and \circled{3} can be formulated by following equations:
\begin{equation}
z^{l+1}_t(FP32) = f_2\times \mathrm{GEMM}(s^{l+1}_t,(B^{l+1}(Ternary))^T) \label{equation12},
\end{equation}
\begin{equation}
f_2 = max|B^{l+1}(FP32)_{max}|,\quad B^{l+1}(Ternary) \in \{-1, 0, +1\} \label{equation13},
\end{equation}
\begin{equation}
m_t^{l+1}(FP32) {=} f_2{\times}f_3 {\times}(\mathrm{GEMM}(e^{l+1}_t(INTX),B^{l+1}(Ternary))) \label{equation14},
\end{equation}
where $f_3$ and $e^{l+1}_t(INTX)$ are calculated in the same manner as $f_1$ in (\ref{equation11}) and $W^l(INTX)$ in (\ref{equation9}), respectively. Note that GEMM \circled{1}~($C_{out}{*}H{*}W{*}C_{in}{*}K_h{*}K_w$) dominates the overall computation cost among the three GEMM operations, and thus we focus on minimizing its bit-width while maintaining training accuracy.

Moreover, to avoid the hardware-unfriendly sWS mentioned in Section~\ref{section_2_2}, we propose using the scaled weight centralization through time~(sWCTT)~\cite{chen2026spikon} to achieve the same normalization effect as sWS. The sWCTT can be described as follows:
\begin{equation}
    \overline{w}^l_{t,i,j}=w^l_{t,i,j}-\mu_{w^l_{t,i,.}},\hat{w}^l_t=\alpha^l_t\overline{w}^l_{t}, w^{l}_{t}=w^{l}\ \forall t \label{equation15}, 
\end{equation}
where $w^l_t$, $\overline{w}^l_{t}$, and $\hat{w}^l_t$ denote the original, centralized, and scaled weights, respectively. $\alpha_t^l$ is a timestep-wise learnable parameters that rescales the centralized weights $\overline{w}_t^l$. Compared to sWS, sWCTT avoids computing variance and simplifies gradient computation.

\begin{algorithm}[tb]
    \caption{One training iteration of Lonic for layer $l$}
    \label{algorithm1}
    \renewcommand{\algorithmicrequire}{\textbf{Input:}}
    \renewcommand{\algorithmicensure}{\textbf{Output:}}
    \begin{algorithmic}[1]
        \REQUIRE Network parameters $w^l$, $\alpha^l_t$; SNN timestep $T$; Learning rate $\eta$; Projection matrix $B^l$; Other required training parameters. 
        \STATE \textbf{Initialize:} $\nabla w^l$=0. 
        \FOR{$t=0,1,\dots, T{-}1$}
            \STATE \textit{Forward:} Calculate $u^{l+1}_t$, $s^{l+1}_t$, and $v^{l+1}_t$ using equation (\ref{equation1}), (\ref{equation2}), (\ref{equation3}), (\ref{equation9}), (\ref{equation10}), (\ref{equation11}) (weight $w^l$ is rescaled and centralized by equation (\ref{equation15}));
            \STATE \textit{Backward:} Calculate instantaneous $\nabla w^l_t$ and $\nabla \alpha^l_t$ by equation (\ref{equation4})-(\ref{equation7}), (\ref{equation12})-(\ref{equation14}), and the corresponding gradient of (\ref{equation15});
            \STATE \textit{Weight gradient accumulation:} $\nabla w^l$=$\nabla w^l$ + $\nabla w^l_t$;
        \ENDFOR
        \STATE \textit{Parameter update:} $w^l$ = $w^l$ $-$ $\eta$$\nabla w^l$, $\alpha^l_t$ = $\alpha^l_t$ $-$ $\eta$$\nabla \alpha^l_t$;
         \ENSURE Trained SNN parameters $w^l$, $\alpha^l_t$.
    \end{algorithmic}
\end{algorithm}

\subsection{Overall Pseudo-code Implementation}

Algorithm~\ref{algorithm1} describes the complete training procedure for a single SNN layer under the fully local online learning paradigm. At each timestep, forward computation updates neuronal states and generates spike outputs based on the low-precision weights~(Line 3). The backward process then computes instantaneous gradients for $\nabla w^l_t$ and $\nabla \alpha^l_t$ using locally available information~(Line 4). Finally, model parameters are updated after processing all timesteps in a single training iteration~(Line 5 and 7). 

\subsection{Hardware-friendly Properties}
\label{section_3_2}
The proposed low-precision training flow brings multiple key advantages for energy-efficient SNN learning hardware implementation. \textit{First,} it allows integer processing elements~(PEs) to replace FP PEs, leading to significant reductions in hardware area and power consumption~(Section~\ref{section_4_2}). \textit{Second,} it introduces additional sparsity beyond the inherent spike sparsity~(Section~\ref{section_4_3}). For example, in GEMM \circled{1}, the quantized $W^l(INTX)$ contains more zero values compared to its FP counterpart $W^l(FP32)$. \textit{Third,} it effectively reduces off-chip weight movement during training~(Section~\ref{section_4_5}). Since integer weights are directly used in GEMM \circled{1}, updated weights can be quantized on-chip before being written back to off-chip memory. This converts FP32 weight transfers into low-precision integer movement, thus reducing off-chip memory access. 

\section{Lonic Hardware Architecture}
\label{section_4}
\subsection{Overview of Lonic Architecture}
\label{section_4_1}
\begin{figure}[tb]
    \centering
    \setlength{\abovecaptionskip}{0pt}
    \includegraphics[width=\linewidth]{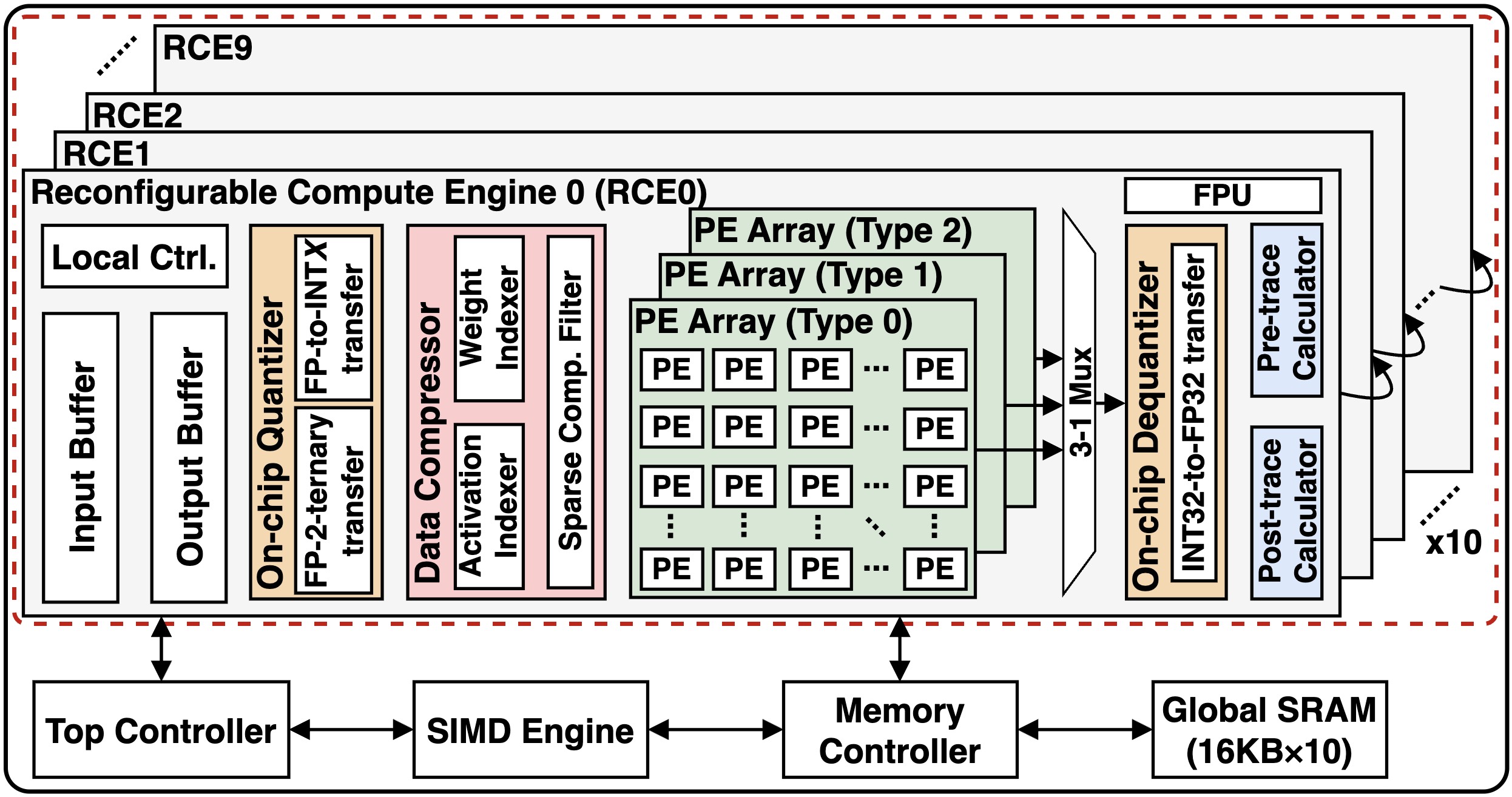}
    \caption{Lonic training accelerator's overall architecture. The sizes of PE array~(type 0/1/2) are $8{\times}8$, $4{\times}4$, and $8{\times}8$ PEs.}
    \Description{fig4}
    \label{fig4}
    \vspace{-10pt}
\end{figure}

To translate algorithmic benefits into practical hardware efficiency, we design the specialized accelerator that enables energy-efficient and scalable fully local online supervised learning for SNNs. As shown in Fig.~\ref{fig4}, Lonic consists of 10 reconfigurable compute engines~(RCEs), a 160-KB global SRAM, a memory controller, a single instruction multiple data~(SIMD) engine, and a top controller. Each RCE is responsible for one timestep in fully local online learning, and the 10 RCEs together enable simultaneous forward, backward, and weight update across all timesteps. The main components of the RCE are the on-chip quantizer~(OCQ), data compressor~(DCOM), PE arrays~(type 0/1/2), on-chip dequantizer~(OCDQ), pre-trace/post-trace calculators, and FP unit. OCQ, PE arrays, and OCDQ are utilized to support the proposed low-precision training flow in Fig.~\ref{fig3}. DCOM leverages spike sparsity and weight sparsity introduced in GEMM \circled{1} to achieve dual-optimization zero gating~(Section~\ref{section_4_3}). Moreover, the SIMD engine performs element- and vector-wise operations, such as max pooling, softmax, and LIF layers.

\begin{figure*}[tb]
    \centering
    \setlength{\abovecaptionskip}{0pt}
    \includegraphics[width=0.85\linewidth]{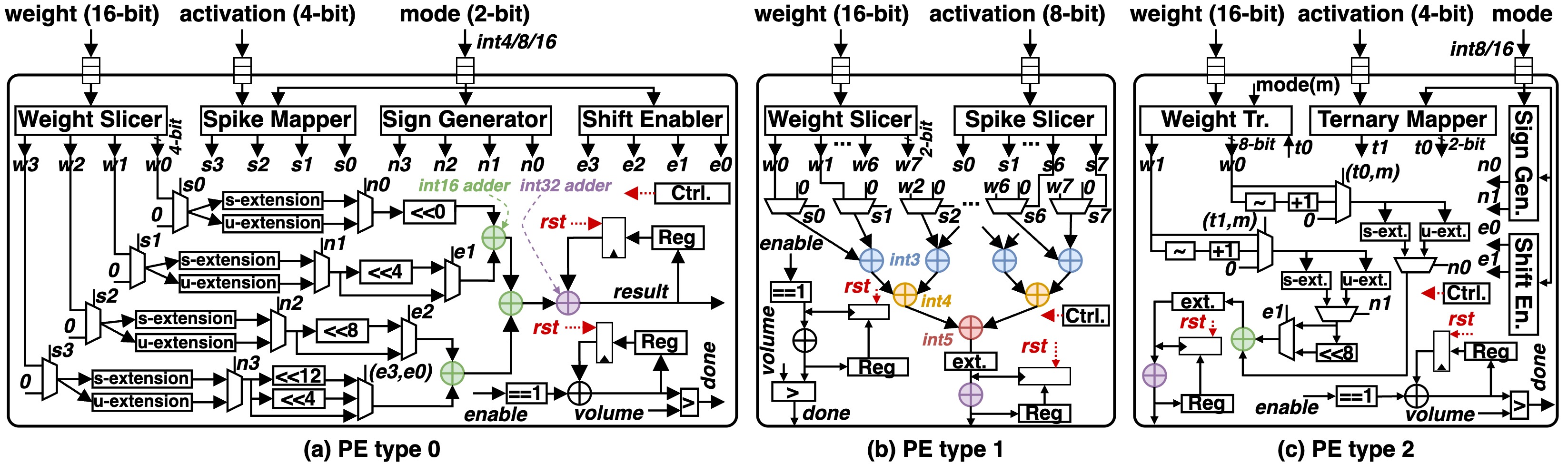}
    \caption{Reconfigurable integer PE design. (a) PE type 0 that supports INT4/8/16. (b) Binary-ternary PE type 1. (c) Ternary-INT8/16 PE type 2. Ctrl.-Controller, Tr.-Transformer, Gen.-Generator, and En.-Enabler.}
    \Description{fig5}
    \label{fig5}
    \vspace{-6pt}
\end{figure*}

\subsection{Reconfigurable Integer PE Design}
\label{section_4_2}
We implement three types of reconfigurable multiplier-free integer PE~(Fig.~\ref{fig5}) for GEMM \circled{1}/\circled{2}/\circled{3} to leverage the benefits of the proposed low-precision training flow. 

\underline{\textit{PE type 0.}} It supports configurable weight precision~(INT4/8/16) through a mode-controlled bit-sliced architecture, enabling flexible trade-offs between accuracy and efficiency. The 16-bit weight is decomposed into four 4-bit segments~($w_0{-}w_3$) by the weight slicer. The spike mapper selects the valid activation bits based on the mode~(e.g., all 4 bits are used in INT4 mode). The sign generator and shift enabler further adapt to the selected mode by determining the signed/unsigned extension and shift bit-width of $w_0/w_1/w_2/w_3$. For example, in INT16 mode, only $w_3$ is sign-extended, while $w_1$, $w_2$, and $w_3$ are left-shifted by 4, 8, and 12 bits, respectively. Three INT16 adders and one INT32 adder are utilized to complete the accumulation in Fig.~\ref{fig5}(a). The partial result is accumulated over multiple cycles, and the $done$ signal is asserted once the processed data reaches the preset $volume$. Note that \textit{PE type 0} achieves a throughput ratio of $4{:}2{:}1$ under INT4, INT8, and INT16 modes. 

\underline{\textit{PE type 1.}} It is designed to efficiently complete the binary-ternary GEMM \circled{2} in Fig.~\ref{fig3}. Due to the small output feature map size~($N$), the processing parallelism is set to 8 to accelerate each output element's computation. Specifically, 16-bit weights~(ternary $B^l$) are decomposed into eight 2-bit segments, while activations~(spikes) are sliced into eight 1-bit segments. The signals $s_0{-}s_7$ determine whether $w_0{-}w_7$ contribute to the computation via multiplexers, thus skipping zero operations. To reduce hardware overhead, a lightweight three-level adder tree is adopted, consisting of INT3, INT4, and INT5 adders at each level, respectively. The INT32 adder and $done$ signal serve the same function as those in \textit{PE type 0}. 

\underline{\textit{PE type 2.}} It supports two modes for the ternary-INT8/16 GEMM \circled{3} in Fig.~\ref{fig3}. Different from \textit{PE type 0}, the input activations are two ternary segments, where the value $-1$ requires sign inversion of the corresponding weights. In INT8 mode, $t_0$ and $t_1$ determine whether the $w_0$ and $w_1$ should be negated. In contrast, in INT16 mode, the original 16-bit weight is first negated through the weight transformer before being sliced into $w_0$ and $w_1$, since independently negating $w_0$/$w_1$ leads to incorrect results.

\begin{figure}[tb]
    \centering
    \setlength{\abovecaptionskip}{0pt}
    \includegraphics[width=\linewidth]{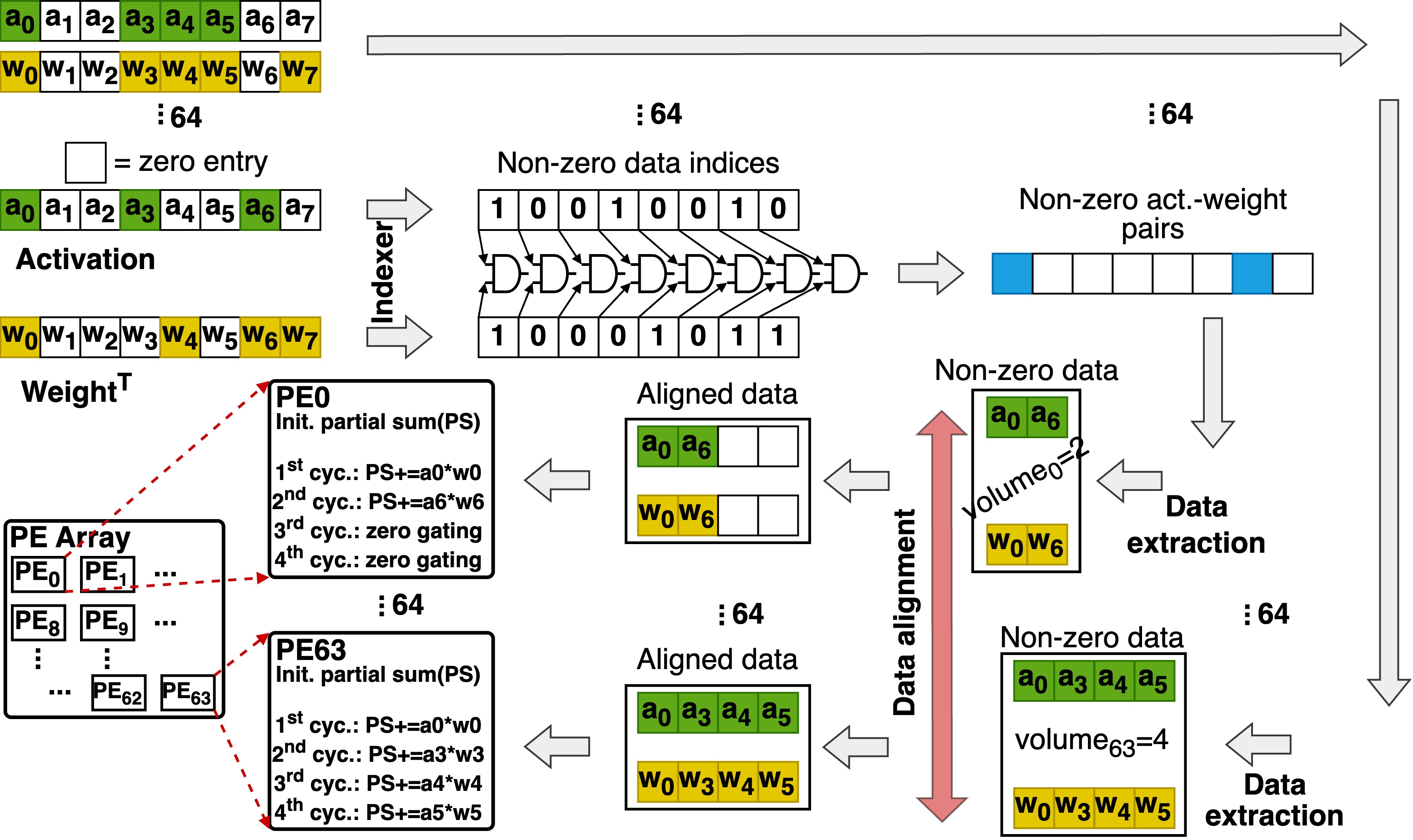}
    \caption{Dual-optimization zero-gating method.}
    \Description{fig6}
    \label{fig6}
    \vspace{-12pt}
\end{figure}

\subsection{Dual-optimization Zero-gating Strategy}
\label{section_4_3}
To leverage the benefits of our low-precision training flow, which introduces additional sparsity~(Section~\ref{section_3_2}), we are inspired to use spike and weight sparsity to reduce energy consumption and improve throughput simultaneously. Prior works~\cite{chen2016eyeriss, yin2024loas, chen2025titanus} exploit activation~(act.) sparsity or joint act.-weight sparsity to skip redundant computations, reducing energy consumption. NVIDIA~\cite{mishra2021accelerating} further proposes a structured $2{:}4$ sparsity pattern to improve both throughput and energy efficiency. In contrast, we aim to fully exploit unstructured sparsity in weights and spikes, translating algorithmic advantages into hardware efficiency through a dual-optimization zero-gating strategy~(Fig.~\ref{fig6}). Specifically, we design a data compressor~(DCOM)~(Fig.~\ref{fig4}) that processes spikes and integer weights to extract and forward only non-zero act.-weight pairs to the PE array. As shown in Fig.~\ref{fig6}, act. and weight indexers inside DCOM first generate binary masks indicating non-zero data entries. These masks are combined via bit-wise AND operations to identify valid act.-weight pairs. Based on the resulting mask, only non-zero pairs are extracted from the original inputs. For example, the pairs $(a_0, w_0)$ and $(a_6, w_6)$ are selected, yielding $volume{=}2$ for PE0. Since the PE array contains 64 PEs, a data alignment step is introduced after data extraction to synchronize inputs across all PEs, enabling simultaneous completion of computation and reducing control complexity. PEs with smaller $volumes$ are zero-padded to match the maximum $volume$, and the aligned data are processed in parallel by 64 PEs. During execution, underutilized PEs apply zero-gating in later cycles, while fully loaded PEs remain active. For instance, PE0 performs computation in the first two cycles followed by zero-gating, whereas PE63 remains active across all cycles. 

\begin{figure}[tb]
    \centering
    \setlength{\abovecaptionskip}{0pt}
    \includegraphics[width=\linewidth]{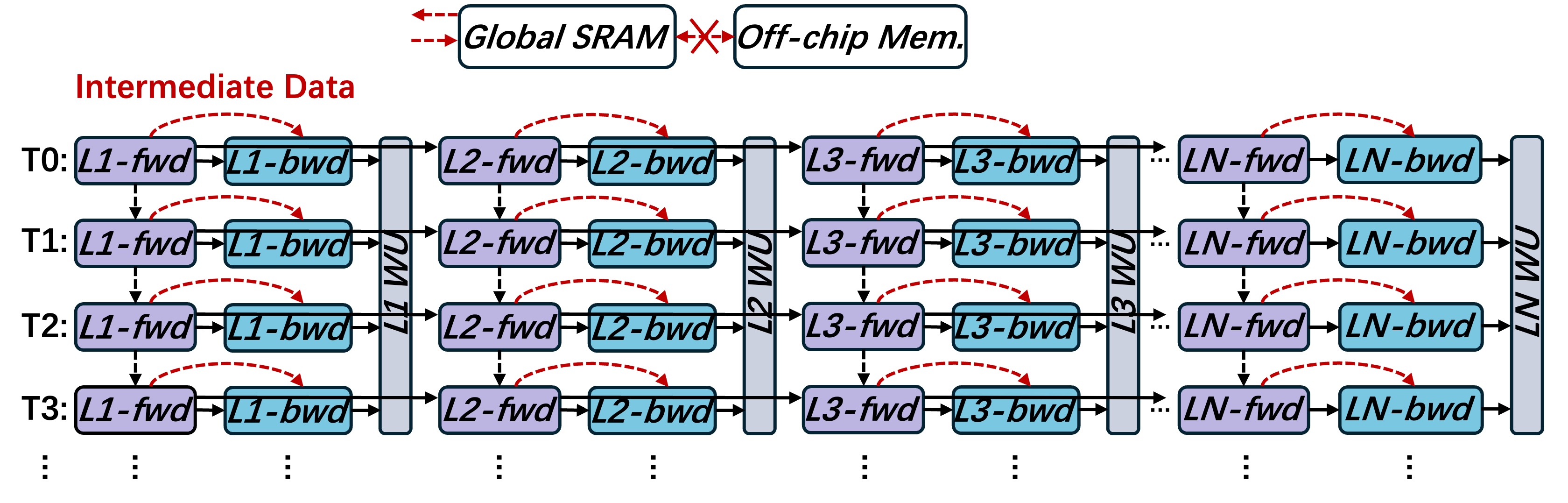}
    \caption{Illustration of the proposed temporal prefixed-accelerated local SNN learning dataflow. FWD-Forward, BWD-Backward, and WU-Weight update.}
    \Description{fig7}
    \label{fig7}
    \vspace{-16pt}
\end{figure}

\begin{table*}[tb]
\centering
\caption{Performance comparison of different temporally and fully local online learning algorithms on CIFAR-10, CIFAR-100, DVS-CIFAR10, and DVS128-Gesture using NVIDIA GPU. ``--'' indicates results that are not reported, as GPU cannot exploit the hardware-level advantages of the proposed low-precision training method. GEMM \protect\circled{1}: $INT4/8/16{-}Spike$, GEMM \protect\circled{2}: $Spike{-}Ternary$, and GEMM \protect\circled{3}: $Ternary{-}INT8$.}
\vspace{-8pt}
\label{table1}
\renewcommand{\arraystretch}{0.85}
\resizebox{\linewidth}{!}{
\begin{tabular}{c|c|c|c|c|c|c|c|c}
\toprule
\textbf{Dataset} & \textbf{Method} & \textbf{Network} & \textbf{Params} & \textbf{Timesteps} & \textbf{Local Learning} & \textbf{Accuracy} & \textbf{Training Time/Epoch (s)} & \textbf{Training Energy/Epoch (J)} \\
\midrule

\multirow{7}{*}{CIFAR-10}
& OTTT  & VGG11 (sWS) & 9.23M & 6  & Partial (in time) & 92.47\% & 2,829.10 & 525,395.45 \\
& SLTT  & VGG11 (sWS) & 9.23M & 6  & Partial (in time) & 92.47\% & 2,048.03 & 397,490.18 \\
& TESS  & VGG11 (sWS) & 9.23M & 6  & Yes & 89.49\% & 2,186.60 & 447,864.42 \\
& \cellcolor{1}Lonic & \cellcolor{1}VGG11 (sWCTT) & \cellcolor{1}9.22M & \cellcolor{1}6  & \cellcolor{1}Yes & \cellcolor{1}88.55\% & \cellcolor{1}1,826.20 & \cellcolor{1}336,972.94 \\
& \cellcolor{1}Lonic & \cellcolor{1}VGG11 (sWCTT+INT4 training) & \cellcolor{1}9.22M & \cellcolor{1}6  & \cellcolor{1}Yes & \cellcolor{1}88.80\% & \cellcolor{1}-- & \cellcolor{1}-- \\
& \cellcolor{1}Lonic & \cellcolor{1}VGG11 (sWCTT+INT8 training) & \cellcolor{1}9.22M & \cellcolor{1}6  & \cellcolor{1}Yes & \cellcolor{1}87.80\% & \cellcolor{1}-- & \cellcolor{1}-- \\
& \cellcolor{1}Lonic & \cellcolor{1}VGG11 (sWCTT+INT16 training) & \cellcolor{1}9.22M & \cellcolor{1}6  & \cellcolor{1}Yes & \cellcolor{1}89.36\% & \cellcolor{1}-- & \cellcolor{1}-- \\

\midrule

\multirow{7}{*}{CIFAR-100}
& OTTT  & VGG11 (sWS) & 9.27M & 6  & Partial (in time) & 70.16\% & 2,886.71 & 524,567.27 \\
& SLTT  & VGG11 (sWS) & 9.27M & 6  & Partial (in time) & 70.29\% & 2,095.40 & 401,829.07 \\
& TESS  & VGG11 (sWS) & 9.27M & 6  & Yes & 64.64\% & 2,195.90 & 447,479.07 \\
& \cellcolor{1}Lonic & \cellcolor{1}VGG11 (sWCTT) & \cellcolor{1}9.27M & \cellcolor{1}6  & \cellcolor{1}Yes & \cellcolor{1}63.62\% & \cellcolor{1}1,783.50 & \cellcolor{1}333,700.62 \\
& \cellcolor{1}Lonic & \cellcolor{1}VGG11 (sWCTT+INT4 training) & \cellcolor{1}9.27M & \cellcolor{1}6  & \cellcolor{1}Yes & \cellcolor{1}62.66\% & \cellcolor{1}-- & \cellcolor{1}-- \\
& \cellcolor{1}Lonic & \cellcolor{1}VGG11 (sWCTT+INT8 training) & \cellcolor{1}9.27M & \cellcolor{1}6  & \cellcolor{1}Yes & \cellcolor{1}63.26\% & \cellcolor{1}-- & \cellcolor{1}-- \\
& \cellcolor{1}Lonic & \cellcolor{1}VGG11 (sWCTT+INT16 training) & \cellcolor{1}9.27M & \cellcolor{1}6  & \cellcolor{1}Yes & 6\cellcolor{1}3.10\% & \cellcolor{1}-- & \cellcolor{1}-- \\

\midrule

\multirow{7}{*}{DVS-CIFAR10}
& OTTT  & VGG11 (sWS) & 9.23M & 10 & Partial (in time) & 76.80\% & 962.6 & 180,017.09 \\
& SLTT  & VGG11 (sWS) & 9.23M & 10 & Partial (in time) & 78.70\% & 707.94 & 128,350.70 \\
& TESS  & VGG11 (sWS) & 9.23M & 10 & Yes & 75.60\% & 720.68 & 139,658.86 \\
& \cellcolor{1}Lonic & \cellcolor{1}VGG11 (sWCTT) & \cellcolor{1}9.22M & \cellcolor{1}10 & \cellcolor{1}Yes & \cellcolor{1}75.90\% & \cellcolor{1}607.19 & \cellcolor{1}111,159.12 \\
& \cellcolor{1}Lonic & \cellcolor{1}VGG11 (sWCTT+INT4 training) & \cellcolor{1}9.22M & \cellcolor{1}10 & \cellcolor{1}Yes & \cellcolor{1}75.60\% & \cellcolor{1}-- & \cellcolor{1}-- \\
& \cellcolor{1}Lonic & \cellcolor{1}VGG11 (sWCTT+INT8 training) & \cellcolor{1}9.22M & \cellcolor{1}10 & \cellcolor{1}Yes & \cellcolor{1}75.20\% & \cellcolor{1}-- & \cellcolor{1}-- \\
& \cellcolor{1}Lonic & \cellcolor{1}VGG11 (sWCTT+INT16 training) & \cellcolor{1}9.22M & \cellcolor{1}10 & \cellcolor{1}Yes & \cellcolor{1}75.80\% & \cellcolor{1}-- & \cellcolor{1}-- \\

\midrule

\multirow{12}{*}{DVS128-Gesture}
& OTTT  & VGG11 (sWS) & 9.23M & 20  & Partial (in time) & 97.22\% & 348.15 & 72,046.33 \\
& SLTT  & VGG11 (sWS) & 9.23M & 20 (10) & Partial (in time) & 97.92\% (96.88\%) & 208.54 (120.74) & 47,133.58 (25,468.13) \\

& TESS  & VGG11 (sWS) & 9.23M & 20 (10) & Yes & 97.22\% (95.14\%) & 240.15 (123.30) & 56,784.64 (29,173.64) \\

& \cellcolor{1}Lonic & \cellcolor{1}VGG11 (sWCTT) & \cellcolor{1}9.22M & \cellcolor{1}20 (10) & \cellcolor{1}Yes & \cellcolor{1}96.53\% (95.49\%) & \cellcolor{1}198.73 (104.68) & \cellcolor{1}46,533.55 (21,493.76) \\

& \cellcolor{1}Lonic & \cellcolor{1}VGG11 (sWCTT+INT4 training) & \cellcolor{1}9.22M & \cellcolor{1}20 (10) & \cellcolor{1}Yes & \cellcolor{1}96.18\% (95.49\%) & \cellcolor{1}-- & \cellcolor{1}-- \\

& \cellcolor{1}Lonic & \cellcolor{1}VGG11 (sWCTT+INT8 training) & \cellcolor{1}9.22M & \cellcolor{1}20 (10) & \cellcolor{1}Yes & \cellcolor{1}96.88\% (95.14\%) & \cellcolor{1}-- & \cellcolor{1}-- \\

& \cellcolor{1}Lonic & \cellcolor{1}VGG11 (sWCTT+INT16 training) & \cellcolor{1}9.22M & \cellcolor{1}20 (10) & \cellcolor{1}Yes & \cellcolor{1}96.88\% (95.14\%) & \cellcolor{1}-- & \cellcolor{1}-- \\

\bottomrule
\end{tabular}}
\vspace{-6pt}
\end{table*}

\subsection{Temporal Prefix-accelerated Dataflow}
\label{section_4_4}
Leveraging the characteristics of fully local online SNN learning, we design a temporal prefix-accelerated~(TPA) local training dataflow~(Fig.~\ref{fig7}) to improve throughput and reduce off-chip data movement~(or on-chip memory pressure). Unlike conventional training flow that processes each timestep's forward~(FWD), backward~(BWD), and weight update~(WU) in order, the TPA dataflow enables simultaneous FWD, BWD, and WU across all timesteps for single SNN layer. As shown in Fig.~\ref{fig7}, since all timesteps' inputs to the first layer are available beforehand during training, the FWD computation can be parallelized across all timesteps~(\textit{prefix-accelerated}). Although temporal dependencies exist between timesteps, they only involve the $\beta v^{l+1}_{t-1}$ term in equation~\ref{equation1}, which can be efficiently computed by the FPU module within RCE. Therefore, the induced delay is negligible and does not hinder the simultaneous execution across all timesteps. Moreover, the BWD computation across timesteps relies only on local learning signals, enabling it to be performed independently after the FWD pass. After all timesteps complete the BWD process for the current layer, the WU is executed on the SIMD engine using the accumulated partial weight gradients. Note that intermediate data generated during the FWD pass can be directly reused by the subsequent BWD pass. As a result, they are buffered in global SRAM without off-chip write-back, and are overwritten after reuse by the next layer's intermediate data. 



\subsection{Off-chip Data Movement Reduction via INT4 Weight Transfer}
\label{section_4_5}

\begin{figure}[tb]
    \centering
    \setlength{\abovecaptionskip}{0pt}
    \includegraphics[width=\linewidth]{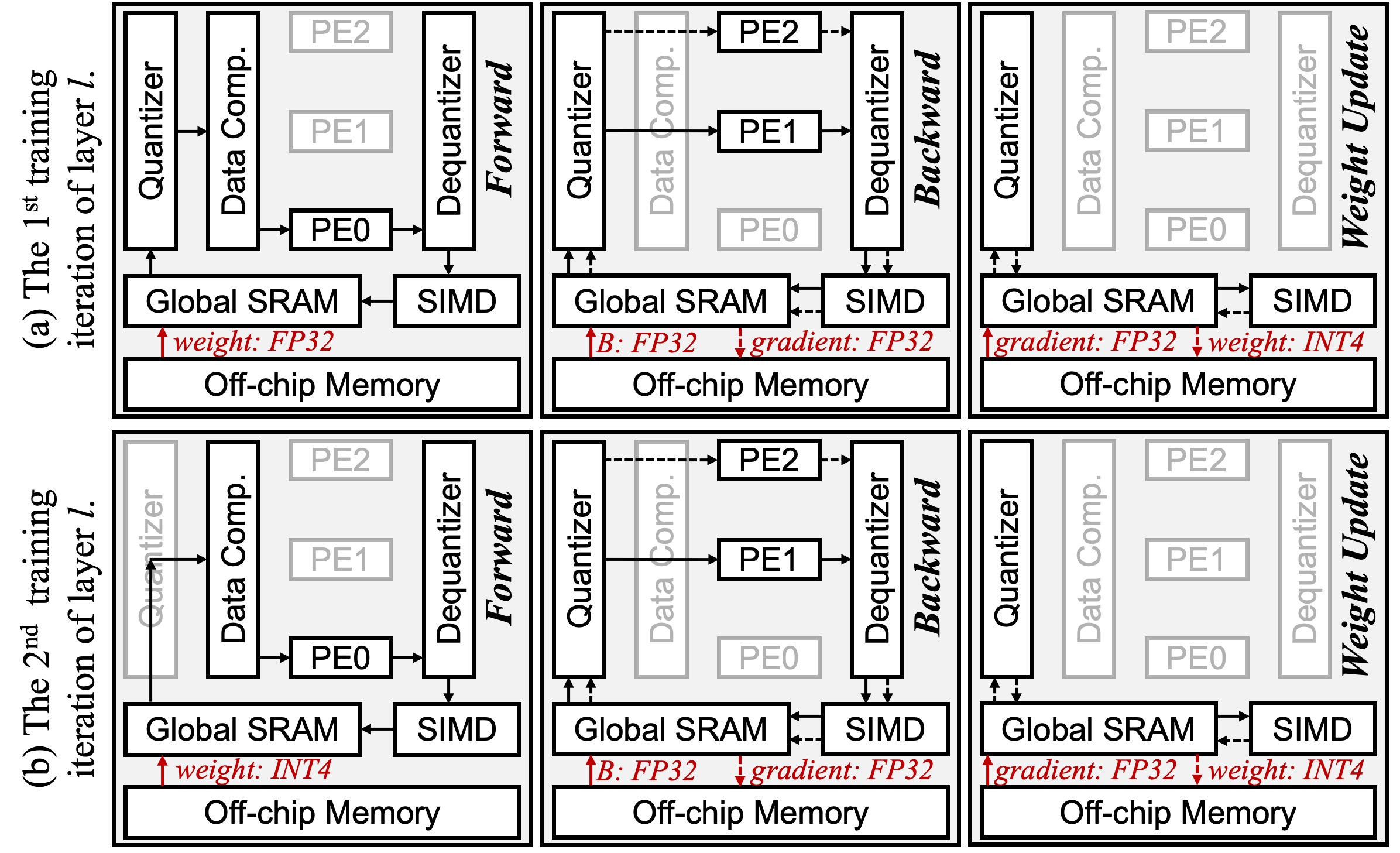}
    \caption {Data movement reduction via low-precision weight transfer during training~(FWD, BWD, and WU).}
    \Description{fig8}
    \label{fig8}
    \vspace{-10pt}
\end{figure}

To further translate algorithmic advantages into hardware efficiency, we propose a low-precision weight transfer scheme that reduces off-chip data movement during training. Fig.~\ref{fig8} illustrates the workflow in the Lonic accelerator during the $1^{st}$ and $2^{nd}$ training iterations for a given layer. \underline{At the $1^{st}$ iteration,} the initialized FP32 weights are loaded into global SRAM. During the FWD pass, they are quantized, compressed by DCOM, and processed by \textit{PE array~(type 0)} for GEMM \circled{1}, followed by dequantization and LIF computation on the SIMD engine. During the BWD pass, the $B$ matrix is first loaded on-chip. In the first round~(solid line), quantized data are processed by \textit{PE array~(type 1)} to perform GEMM \circled{2}, followed by cross-entropy computation on the SIMD engine. In the second round~(dashed line), \textit{PE array~(type 2)} performs GEMM \circled{3}, and the SIMD engine computes weight gradients for the current timestep, which are then written back to off-chip memory. During WU, partial gradients from all timesteps are aggregated on the SIMD engine to update the weights. In the second round, the updated weights are quantized into low-precision integers and written back to off-chip memory. \underline{At the $2^{nd}$ iteration,} different from the $1^{st}$ iteration, only low-precision integer weights are loaded during the FWD pass instead of FP32 weights, while other operations remain unchanged. Therefore, this scheme enables all subsequent iterations to use low-precision weight transfer for loading and write-back during FWD and WU, reducing off-chip memory access.

%% file: sec4_experiment.tex
\section{Experimental Results and Analysis}
\label{section_5}

\begin{figure}[tb]
    \centering
    \setlength{\abovecaptionskip}{0pt}
    \includegraphics[width=\linewidth]{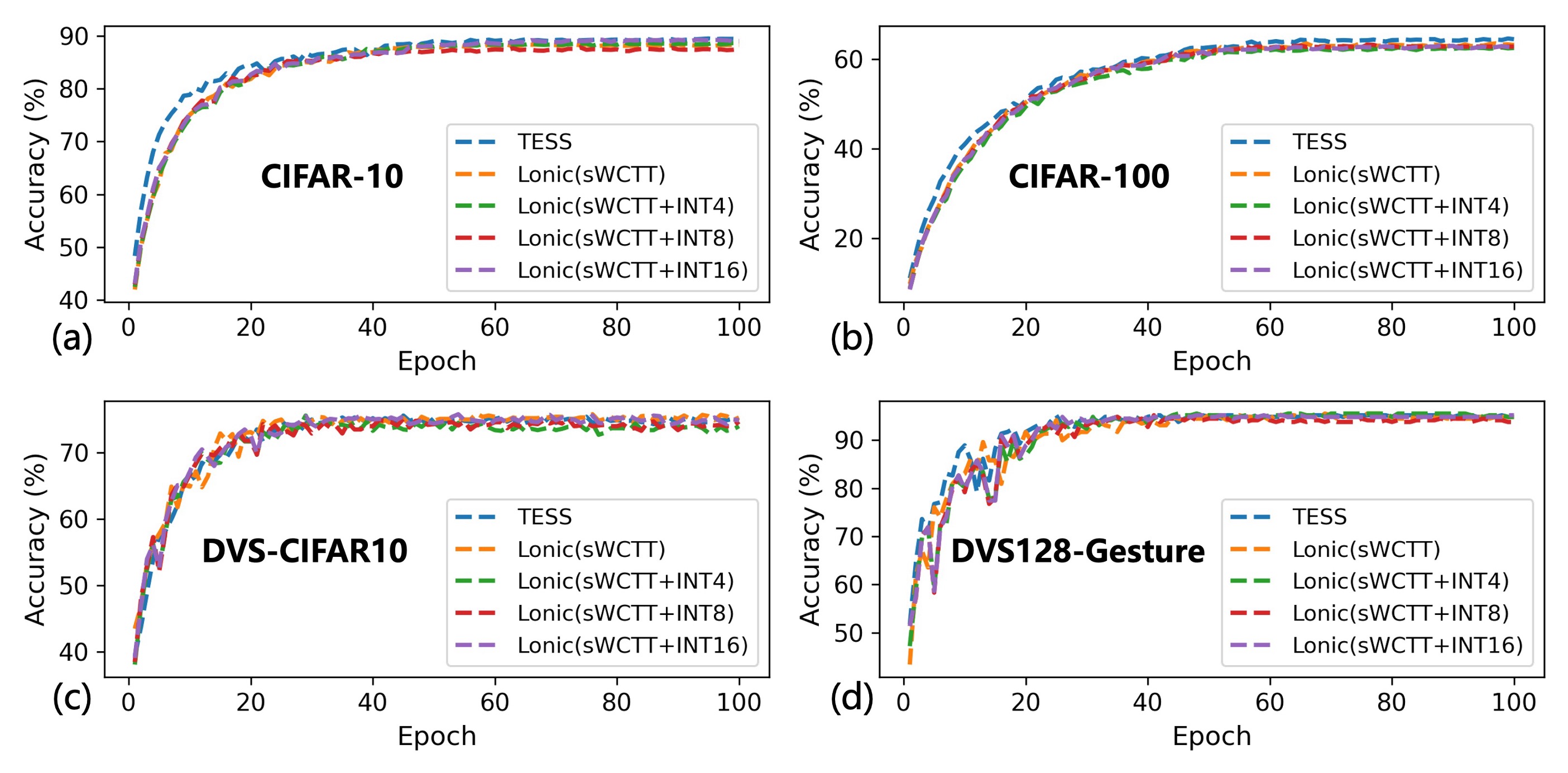}
    \caption {Test accuracy of baseline method and Lonic algorithm under different precision settings. (a) CIFAR-10. (b) CIFAR-100. (c) DVS-CIFAR10. (d) DVS128-Gesture.}
    \Description{fig9}
    \label{fig9}
    \vspace{-10pt}
\end{figure}

\begin{figure}[tb]
    \centering
    \setlength{\abovecaptionskip}{0pt}
    \includegraphics[width=0.8\linewidth]{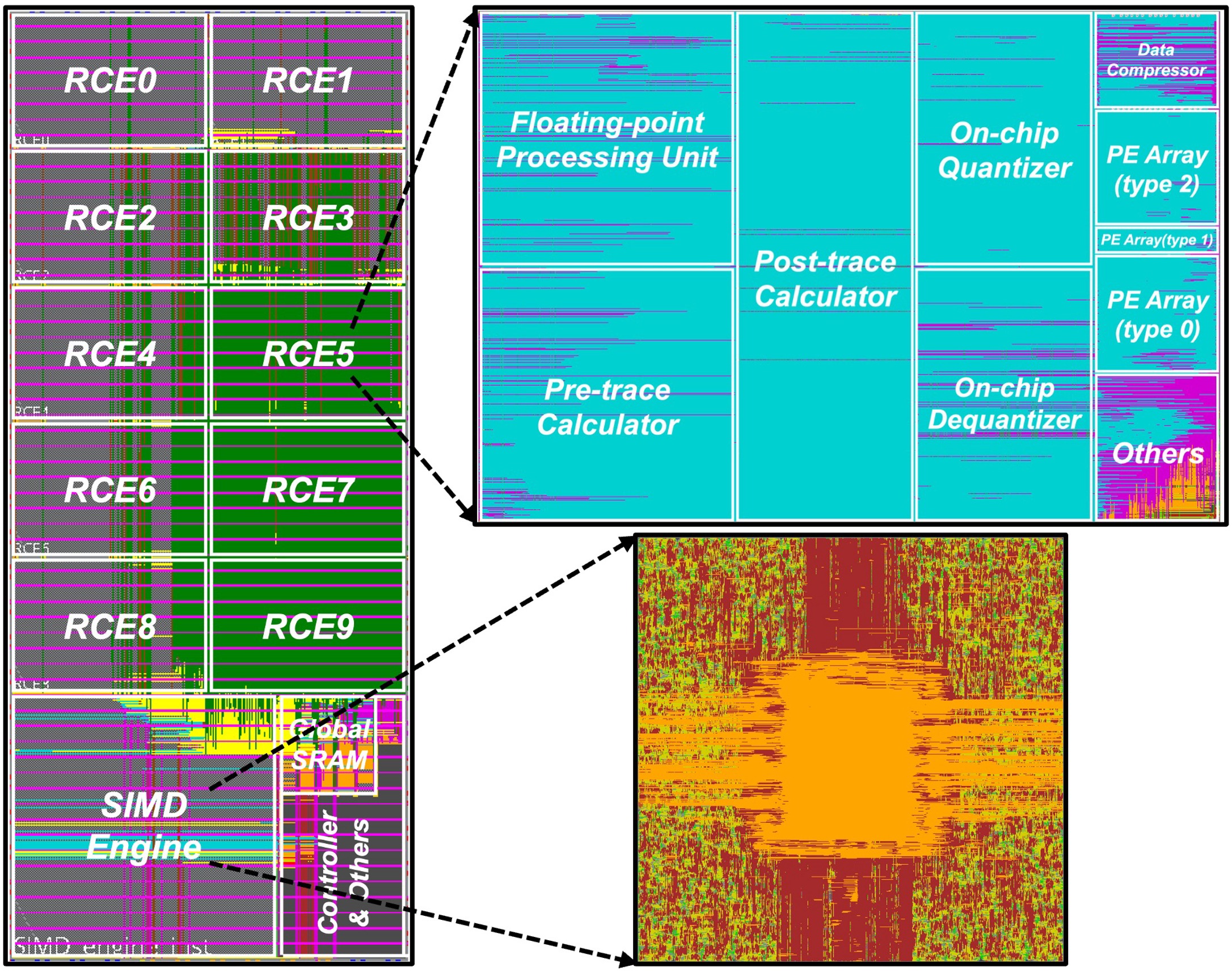}
    \caption {The Layout of our 28nm Lonic accelerator.}
    \Description{fig10}
    \label{fig10}
    \vspace{-12pt}
\end{figure}

\subsection{Experimental Setup}
\label{section_5_1}
\textbf{Algorithm setup.} We evaluate the proposed Lonic algorithm on CIFAR-10~\cite{krizhevsky2009learning}, CIFAR-100~\cite{krizhevsky2009learning}, CIFAR10-DVS~\cite{li2017cifar10}, and DVS128-Gesture~\cite{amir2017low} to demonstrate its superior performance regarding training overhead and accuracy. We use the VGG11 network architecture with a lightweight classifier for all experiments~\cite{simonyan2014very}. The training epoch and batch size of online SNN learning are set to 100 and 1, respectively. Prior local learning algorithms,including OTTT~\cite{xiao2022online}, SLTT~\cite{meng2023towards}, and TESS~\cite{apolinario2025tess}, are used as baseline. 

\textbf{Hardware setup.} We implement the Lonic training accelerator at the RTL-level using Verilog. We synthesize the Lonic accelerator using Synopsys Design Compiler~\cite{synopsys_dc} to estimate the area and power at 1 GHz under the TSMC 28nm node. The global SRAM is created utilizing the TSMC memory compiler. We place and route using Cadence Innovus~\cite{cadence_innovus} to generate the final layout, as shown in Fig.~\ref{fig10}. The energy consumption of off-chip memory HBM2~(5.7pJ/b) is adopted from~\cite{chatterjee2017architecting}. Moreover, we develop a cycle-accurate simulator to evaluate the Lonic accelerator's performance under different tasks.

\textbf{Baselines.} Due to the lack of SOTA hardware baselines on fully local online SNN learning, we compare the Lonic accelerator with Nvidia V100 GPU, Apple M4 GPU~(10 cores), and TPU-like~($18.18mm^2$@28nm) and H2Learn~\cite{liang2021h2learn}~($110.46mm^2$@28nm) training accelerators. Note that we implement the TPU-like baseline with 1444 FP PEs, where a systolic array replaces the RCEs in~Fig.~\ref{fig4}, and evaluate it under our simulator framework. 

\subsection{Lonic Training Algorithm Performance}
\label{section_5_2}
\textbf{Comparison with TESS baseline.} The proposed Lonic algorithm is designed for energy-efficient fully local online SNN learning rather than optimizing for the best accuracy. However, Lonic~(sWCTT + low precision training) still achieves competitive results compared to the SOTA baseline, as shown in Table~\ref{table1}. Since \textit{PE array (type 0)} supports reconfigurable weight precision, we report the results of Lonic with different configurations \textit{sWCTT+INT4/8/16 training}. \underline{(1) Accuracy.} We observe that Lonic exhibits only marginal accuracy degradation compared to TESS across all cases. Specifically, Lonic~(sWCTT+INT4) exhibits an average accuracy drop of 0.67\%, while Lonic~(sWCTT+INT8) and~(sWCTT+INT16) show average drops of 0.76\% and 0.36\%, respectively. We can flexibly select different training settings based on accuracy requirements. \underline{(2) Latency.} Lonic~(sWCTT) achieves a 16.67\% reduction on average in training latency compared to TESS due to the hardware-friendly weight centralization method from our algorithm. The latency and energy reduction of low-precision settings are not reported here, as GPU cannot translate algorithm-level benefits into real-device efficiency. \underline{(3) Energy.} In terms of training energy, Lonic~(sWCTT) achieves an average reduction of 22.99\% over baseline across all cases.

\textbf{Training convergence of Lonic.} As shown in Fig.~\ref{fig9}, the Lonic algorithm demonstrates similar convergence behavior to the TESS baseline across all datasets. Moreover, different precision configurations~(INT4/8/16) exhibit nearly identical convergence trends, indicating that the proposed low-precision training method does not affect training stability and validates the robustness of Lonic.

\begin{figure}[tb]
    \centering
    \setlength{\abovecaptionskip}{0pt}
    \includegraphics[width=\linewidth]{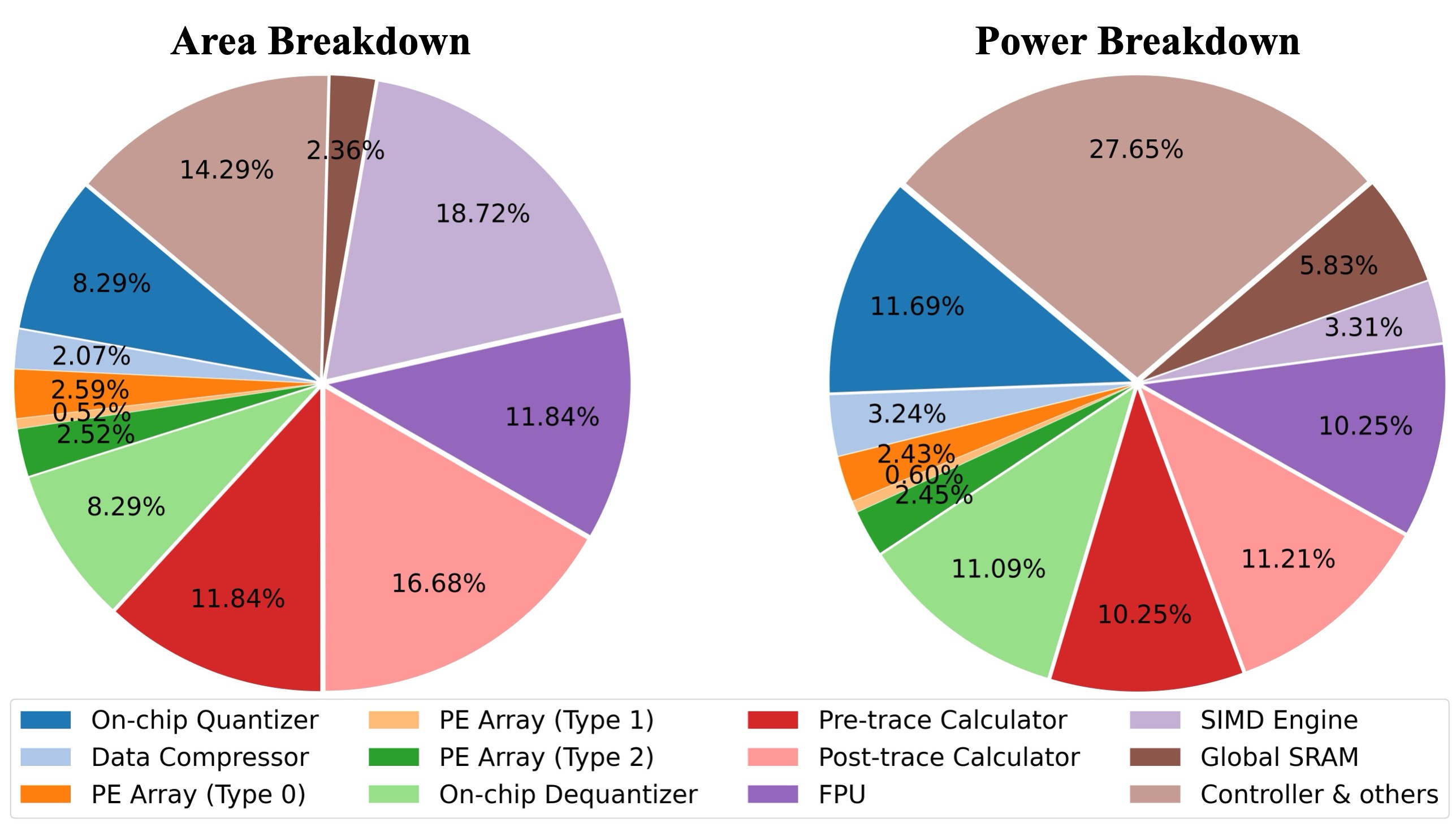}
    \caption {Area and power breakdown of Lonic accelerator.}
    \Description{fig11}
    \label{fig11}
    \vspace{-12pt}
\end{figure}

\begin{figure*}[tb]
    \centering
    \setlength{\abovecaptionskip}{0pt}
    \includegraphics[width=\linewidth]{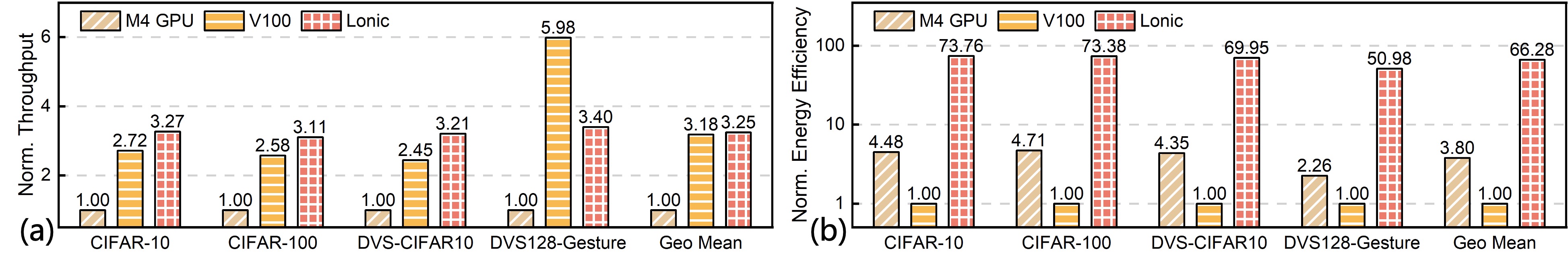}
    \caption{Normalized training throughput~(iteration/s)~(a) and energy efficiency~(iteration/J)~(b) of Lonic accelerator compared to Apple M4 and Nvidia V100 GPUs on both static and neuromorphic datasets.}
    \Description{fig12}
    \label{fig12}
    \vspace{-8pt}
\end{figure*}

\subsection{Lonic Accelerator Evaluation} 

\textbf{Area and power breakdown.} Lonic training accelerator's total area and power are $\mathbf{15.47\,mm^2}$ and $\mathbf{4.05\,W}$, respectively. Fig.~\ref{fig11} presents a detailed area and power breakdown of Lonic. The \textit{PE arrays~(type 0/1/2)} only account for 5.63\% of area and 5.48\% of power, mainly due to the use of integer PEs instead of FP units. To support the proposed dual-optimization zero-gating strategy, the data compressor only incurs 2.07\% in area and 3.24\% in power. To enable efficient on-chip (de)quantization within our low-precision training flow, the quantizer and dequantizer together introduce 15.68\% cost in area and 22.78\% cost in power. Among the remaining modules, FPU is a dedicated on-chip unit with a parallelism of 64 to efficiently perform MAC operations, alleviating the processing pressure on the SIMD engine. In addition, pre-trace and post-trace calculators are necessary components to support gradient computation in fully local online SNN learning. 

\textbf{Comparison with GPUs.} We benchmark the Lonic training accelerator against edge M4~(10 cores) and server V100 GPUs, as shown in Fig.~\ref{fig12}. Note that the Lonic accelerator adopts the proposed low-precision training setting~(sWCTT+INT4), whereas GPUs use the full-precision setting~(sWCTT), since GPUs cannot effectively exploit our Lonic algorithm. We observe that Lonic achieves 3.25x throughput and 17.44x energy efficiency on average across different tasks compared to the M4 GPU. Moreover, Lonic delivers 1.02x and 66.28x improvement in throughput and energy efficiency over the V100 GPU, respectively. The relatively limited throughput gain over V100 GPU stems from the constrained parallelism of the PE arrays within each RCE. This limitation becomes more obvious on the DVS128-Gesture dataset, where the large $128{\times}128$ input feature map leads to substantial GEMM computations. In this scenario, V100 GPU can better leverage its massive parallel processing capability, achieving up to 5.98x speedup, compared to 3.40x for Lonic.

\begin{table}[tb]
\centering
\Large
\caption{Normalized training throughput, energy efficiency, and area efficiency of Lonic co-design compared to TPU-like and H2Learn ASIC accelerators.}
\vspace{-8pt}
\label{table2}
\renewcommand{\arraystretch}{1.2}
\resizebox{\linewidth}{!}{
\begin{tabular}{c|c|c|c}
\toprule
\multirow{2}{*}{\textbf{Algorithm @ Hardware}}
 & \textbf{Norm. Throughput} & \textbf{Norm. Energy Eff.} & \textbf{Norm. Area Eff.} \\
& \textbf{(iteration/s)} & \textbf{(iteration/J)} & \textbf{(iteration/s/mm$^2$)} \\
\midrule
Lonic(sWCTT)@TPU-like 
& 1.00$\times$ 
& 1.00$\times$ 
& 1.00$\times$ \\

Lonic(sWCTT)@H2Learn 
& 12.70$\times$ 
& 10.48$\times$ 
& 2.01$\times$ \\

Lonic(sWCTT+INT4)@Lonic
& 12.46$\times$ 
& 15.95$\times$ 
& 14.64$\times$ \\
\bottomrule
\end{tabular}}
\vspace{-12pt}
\end{table}

\textbf{Comparison with ASIC training accelerators.} Similar to GPUs, TPU-like and H2Learn accelerators also adopt the full-precision training setting~(sWCTT), as their architectures are not designed to support the proposed Lonic algorithm and instead rely on FP units for GEMM computation. \underline{Throughput.} As shown in Table~\ref{table2}, Lonic exhibits 12.46x speedup on average compared to the TPU-like design. However, we observe that Lonic only achieves slight throughput improvements over H2Learn on CIFAR-10, CIFAR-100, and DVS-CIFAR10 datasets. H2Learn also outperforms Lonic in throughput on DVS128-Gesture. This is because H2Learn is designed with a larger chip area~($110.46mm^2$@28nm) and higher computation capability~(27.85TFLOPS, 1.8x of V100 GPU), allowing it to efficiently process the large GEMM workloads. Nevertheless, Lonic, enabled by the proposed optimization methods, still achieves comparable throughput to H2Learn on average. \underline{Energy Efficiency.} Compared to TPU-like and H2Learn, Lonic achieves 15.95x and 1.52x energy efficiency, respectively. \underline{Area Efficiency.} In terms of area efficiency, due to its smaller area footprint, Lonic achieves 14.64x and 7.28x improvements over TPU-like and H2Learn, respectively.

\begin{figure}[tb]
    \centering
    \setlength{\abovecaptionskip}{0pt}
    \includegraphics[width=\linewidth]{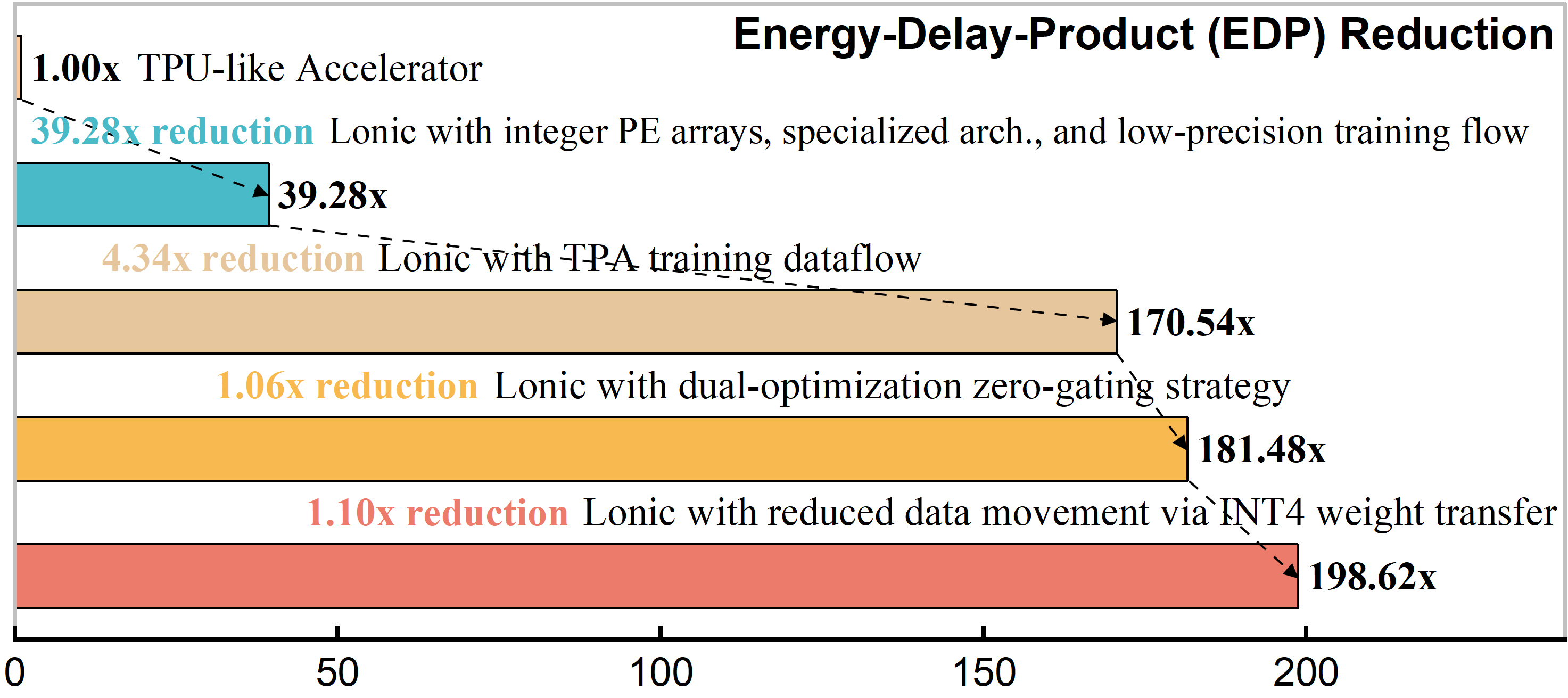}
    \caption {Ablation study of Lonic accelerator.}
    \Description{fig14}
    \label{fig14}
    \vspace{-12pt} 
\end{figure}

\textbf{Ablation study on proposed techniques.} We conduct an ablation study to evaluate the effectiveness of the proposed optimization methods. All results are reported as averages across different training tasks. To jointly consider throughput and energy efficiency, we use the energy-delay product~(EDP) as the evaluation metric. As shown in Fig.~\ref{fig14}, compared to the TPU-like accelerator baseline, Lonic with integer PE arrays, specialized architecture, and low-precision training algorithm achieves a 39.28x EDP reduction. By further introducing the TPA local learning dataflow, the EDP reduction is significantly improved to 170.54x. The proposed dual-optimization zero-gating strategy provides additional gains, especially for the large GEMM workloads in the DVS128-Gesture dataset, increasing the EDP reduction to 181.48x. Finally, by reducing off-chip data movement through INT4 weight transfer, Lonic achieves the highest EDP reduction of 198.62x.

\textbf{Scalability study of Lonic accelerator.} Lonic accelerator can scale to support the training of large SNN models~(e.g., VGGs and ResNets), as the SIMD engine enables efficient execution of various element-wise and vector-wise operations. We benchmark TPU-like, H2Learn, and Lonic accelerators under the VGG16 SNN model with a lightweight classifier across different datasets. We observe that Lonic achieves average speedups of 9.73x and 1.02x over the TPU-like and H2Learn accelerators, respectively. Lonic also outperforms TPU-like and H2Learn in energy efficiency, achieving 12.29x and 1.43x improvements, respectively. Moreover, Lonic delivers 11.44x and 7.58x improvement in area efficiency compared to TPU-like and H2Learn, respectively.

%% file: sec5_conclusion.tex
\section{Conclusion}
\label{section_6}
We present Lonic, an algorithm-hardware co-design framework for energy-efficient and scalable fully local online supervised SNN learning. On the algorithm side, we propose an INT4 low-precision training flow for fully local online SNN learning while maintaining the accuracy. On the hardware side, we leverage reconfigurable multiplier-free integer PEs, dual-optimization zero-gating scheme, TPA local learning dataflow, and low-precision weight movement to significantly improve training efficiency. Lonic provides a promising solution for energy-efficient local online SNN training by bridging algorithmic efficiency and hardware acceleration.